\documentstyle[12pt]{article}

\makeatletter
\def\vereq#1#2{\lower3pt\vbox{\baselineskip1.5pt \lineskip1.5pt
\ialign{$\m@th#1\hfill##\hfil$\crcr#2\crcr\sim\crcr}}}
\makeatother

\def\lesssim{\mathrel{\mathpalette\vereq<}}
\def\gtrsim{\mathrel{\mathpalette\vereq>}}

\makeatletter
\def\lesssim{\mathrel{\mathpalette\vereq<}}
\def\vereq#1#2{\lower3pt\vbox{\baselineskip1.5pt \lineskip1.5pt
\ialign{$\m@th#1\hfill##\hfil$\crcr#2\crcr\sim\crcr}}}

\def\gtrsim{\mathrel{\mathpalette\vereq>}}
\makeatother

\newcommand{\Tr}{\mbox{Tr}}
\newcommand{\Det}{\mbox{det}}

\newcommand{\beq}{\begin{equation}}
\newcommand{\eeq}{\end{equation}}

\newcommand{\remove}[1]{}

\begin{document}
\begin{titlepage}
\begin{center}
\today     \hfill    LBL-37099\\


{\large \bf Third family flavor physics \\
in an SU(3)$^{\rm{\bf 3}}\times$SU(2)$_{\rm{\bf L}}
\times$U(1)$_{\rm{\bf Y}}$ model}\footnote{This
work was supported by the Director, Office of Energy Research, Office of
High Energy and Nuclear Physics, Division of High Energy Physics of the
U.S. Department of Energy under Contract DE-AC03-76SF00098.}


\vskip 0.3in

Christopher D. Carone and Hitoshi Murayama\footnote{On leave of absence from
{\it Department of Physics, Tohoku University, Sendai, 980 Japan.}}

{\em Theoretical Physics Group\\
     Lawrence Berkeley Laboratory\\
     University of California\\
     Berkeley, California 94720}

\end{center}

\vskip .3in

\begin{abstract}
We consider a model in which each family transforms under a different
SU(3) color group.  The low-energy effective theory is an extension
of the Standard Model, with additional color octet gauge bosons
$G_H$ with mass $M$ that couple preferentially to the third family quarks.
We show that there are two distinct regions of the model's parameter space
in which we can simultaneously evade all the current experimental
constraints, one with $M \approx 250$ GeV and the other with
$M \gtrsim 600$ GeV.  Within each allowed region, we can obtain a
correction to the $Zb\overline{b}$ vertex that is consistent with the
slightly high value of $R_b$ observed at LEP.  We show that there are
$\Delta B = 1$ operators in our model that can suppress the $B$-meson
semileptonic branching ratio $B_{SL}$ and the charm multiplicity per
decay $n_c$ by enough to reconcile the spectator parton model predictions
with the experimental data.  In the non-supersymmetric version of our
model, we can only obtain the desired corrections to $R_b$, $B_{SL}$
and $n_c$ in different regions of the allowed parameter space, while in
the supersymmetric version, we can obtain all three corrections
simultaneously.  We also discuss a strong-coupling limit of our model
in which the third-family quarks become composite.
\end{abstract}

\end{titlepage}
\renewcommand{\thepage}{\roman{page}}
\setcounter{page}{2}
\mbox{ }

\vskip 1in

\begin{center}
{\bf Disclaimer}
\end{center}

\vskip .2in

\begin{scriptsize}
\begin{quotation}
This document was prepared as an account of work sponsored by the United
States Government. While this document is believed to contain correct
information, neither the United States Government nor any agency
thereof, nor The Regents of the University of California, nor any of their
employees, makes any warranty, express or implied, or assumes any legal
liability or responsibility for the accuracy, completeness, or usefulness
of any information, apparatus, product, or process disclosed, or represents
that its use would not infringe privately owned rights.  Reference herein
to any specific commercial products process, or service by its trade name,
trademark, manufacturer, or otherwise, does not necessarily constitute or
imply its endorsement, recommendation, or favoring by the United States
Government or any agency thereof, or The Regents of the University of
California.  The views and opinions of authors expressed herein do not
necessarily state or reflect those of the United States Government or any
agency thereof, or The Regents of the University of California.
\end{quotation}
\end{scriptsize}

\vskip 2in

\begin{center}
\begin{small}
{\it Lawrence Berkeley Laboratory is an equal opportunity employer.}
\end{small}
\end{center}

\newpage
\renewcommand{\thepage}{\arabic{page}}
\setcounter{page}{1}
\section{Introduction}

Despite its enormous success, the Standard Model gives us no
explanation of the origin of flavor.  The reason for the existence
of three generations of fermions and for their hierarchical pattern of
masses and mixing angles is, at present, not understood.  The flavor
physics of the third generation has been particularly mysterious.
The smallness of the third generation mixing angles, and the huge
hierarchy between the top and bottom quark masses have lead some
to speculate that the third generation may be fundamentally different
from the other two.  Recent experimental results have also helped to
motivate such speculation.  There are a number of small discrepancies
between the current experimental data and the expectations of the
Standard Model concerning properties of the $b$-quark.
The observed $B$-meson semileptonic branching ratio is much lower than the
spectator parton model prediction, and the charm multiplicity in the decay
products is also low, in both cases by roughly 20\% \cite{pdg}.  Another
possible anomaly is the high value of the $Z\rightarrow b\overline{b}$
width observed at LEP, though in this case the discrepancy with the
Standard Model prediction is much smaller. Whether or not these are signs
of new physics or merely systematic errors is subject to
debate \cite{kag,shi}.  Nevertheless, it is natural to question
whether there are models that can explain the observed pattern
of discrepancies, while remaining consistent with all the other relevant
experimental constraints.

In this paper, we consider a model based on the gauge group
SU(3)$_1\times$ SU(3)$_2\times$ SU(3)$_3\times $SU(2)$_L\times$U(1)$_Y$.
Quarks belonging to generation $x$ transform as triplets under SU(3)$_x$,
but as singlets under the other two SU(3)'s. We assume that
SU(3)$_1\times$SU(3)$_2$ breaks to its diagonal subgroup SU(3)$_{1,2}$
at some high scale $\Lambda_H$, and that SU(3)$_{1,2} \times$ SU(3)$_3$
breaks to its diagonal subgroup, ordinary QCD, at a lower scale
$\Lambda_L$.  As we will see, the effective theory below $\Lambda_L$ is
simply the Standard Model with additional massive color octet gauge bosons,
$G_H$, that couple preferentially to the third generation quarks.  We will
show that there are two regions of the parameter space of the model that are
consistent with the constraints from the top quark production cross
section, from collider searches for new particles decaying to dijets, and
from flavor changing neutral current (FCNC) processes.  In the latter
case, we use the approach of Froggatt and Nielsen \cite{fn} to construct
mass matrices that lead to a natural suppression of the neutral flavor
changing $\overline{q} G_H q$ couplings that arise at tree-level in the model.
We show that it is possible within both of these allowed regions to obtain
corrections to the nonleptonic $B$-decay widths that can reconcile the
spectator parton model predictions for the semileptonic branching ratio and
the average charm multiplicity per decay with the measured values.
We also can obtain a correction to the $Z\rightarrow
b\overline{b}$ width that reconciles the Standard Model expectation with the
value observed at LEP, but in a different region of the parameter space.
In the supersymmetric version of our model, however, we can obtain all
three effects simultaneously. As a consequence of the new contribution to the
$Z\rightarrow b\overline{b}$ width, the value of $\alpha_s (m_Z)$
extracted from $\Gamma(Z \rightarrow \mbox{hadrons})$ is reduced.  This shift
is sufficient to reconcile the LEP value of $\alpha_s$ with the measurements
made at lower energies~\cite{shi}.

We should point out that the idea of introducing different gauge groups
for each generation has appeared in various forms throughout the
literature.  However, most of  the early references do not focus on the
detailed phenomenology of the models \cite{early}, while the more recent
references consider different SU(2)$_L$s rather than SU(3)s to explain a
now nonexistent anomaly in $\tau$ decay \cite{ma}.  Our work is also
similar in some respects to the recent literature on the phenomenology of
topcolor models, which also involve additional massive color octet
bosons \cite{hill}.  In contrast to these references, we focus on the
problem of constructing realistic mass matrices and the associated flavor
physics.  Since it is our purpose here to explore the anomalies described
above, rather than to explain electroweak symmetry breaking through top
condensation, we do not restrict ourselves to the case where
the $\overline{q} G_H q$ coupling attains its critical value.  In the
strong-coupling limit of our model, we consider the possibility that the
third generation quarks may become composite, in a way similar to the
Abbott-Farhi model \cite{afar}, without any condensation at all.

The paper is organized as follows.  In Section~\ref{sec:model}, we present
both the non-supersymmetric and supersymmetric versions of our model, and
discuss the specific structure of the fermion mass matrices. The latter
is crucial in determining the constraints from flavor-changing
processes that we present in Sections \ref{sec:param} and \ref{sec:bphys}.
In Section~\ref{sec:param} we identify the allowed parameter space of our
model, and show that we can reproduce the value of $R_b$ measured at LEP
within the two allowed regions.  In Section~\ref{sec:bphys}, we consider
the $\Delta B = 1$ and $\Delta B = 2$ operators that result from
a slightly more complicated form of the down-quark mass matrix.  We show
that the new $\Delta B = 1$ operators can reduce both the $B$-meson
semileptonic branching ratio $B_{SL}$ and the average charm
multiplicity per decay $n_c$ enough to reconcile the experimental
data with the spectator parton model predictions.  We show that
the strong constraints from $b \rightarrow s \gamma$ in the
non-supersymmetric version of the model only allow us to account
for $R_b$ and the $B$ decay anomalies in different regions of the allowed
parameter space.  However, in the supersymmetric version, the
$b \rightarrow s \gamma$ branching fraction is suppressed, allowing
us to explain  $R_b$, $B_{SL}$ and $n_c$ simultaneously.  In
Section~\ref{sec:conf} we consider what may happen in the confining
phase of the color SU(3) corresponding to the third family, and
argue that the bottom and top quarks may become composite.  In
the final section we summarize our conclusions.

\section{The Model} \label{sec:model}

We assume that the color gauge group SU(3)$_1\times$SU(3)$_2\times$SU(3)$_3$
is broken to SU(3)$_{1,2} \times$SU(3)$_3$ at some high scale
$\Lambda_H \gtrsim {\cal O}(100)$ TeV by the vacuum expectation value
(vev) of a Higgs boson $\Phi_{1,2}$ transforming as a ({\bf 3},{\bf
3$^*$},{\bf 1}). In the effective theory below $\Lambda_H$, the first and
second generation fields have color charges under SU(3)$_{1,2}$ and the third
generation fields under SU(3)$_3$. The group SU(3)$_{1,2}\times
$SU(3)$_3$ is broken to its diagonal subgroup, SU(3)$_c$, at a much lower
scale $\Lambda_L$ by the vev of a Higgs boson $\Phi_{2,3}$ transforming as
({\bf 1},{\bf 3},{\bf 3$^*$}) under the original color group. The QCD coupling
$g_s$ is related to the couplings $g_{1,2}$ and $g_3$ by
\begin{equation}
\frac{1}{g_s^2} = \frac{1}{g_{1,2}^2} + \frac{1}{g_{3}^2} ,
\end{equation}
and thus we use the parameterization
\begin{eqnarray}
g_{1,2} &=& \frac{g_s}{\cos \theta},\\
g_{3} &=& \frac{g_s}{\sin \theta} \ .
\end{eqnarray}
The last stage of symmetry breaking leaves a color-octet vector
boson $G_H$ with mass $M \sim g_s \Lambda_L/(\sin\theta\cos\theta)$
and with the following coupling to the quarks:
\begin{equation}
{\cal L} = g_s G_H^{\mu a}
      (\bar{u}', \bar{c}', \bar{t}') \gamma_\mu T^a
      \left( \begin{array}{ccc}
            \tan \theta & 0 & 0\\
            0 & \tan \theta & 0\\
            0 & 0 & - \cot \theta
             \end{array} \right)
      \left( \begin{array}{c}
            u' \\ c' \\ t'
             \end{array} \right) + \mbox{down-sector} \ .
\end{equation}
Here, the $T^a$ are $SU(3)$ generators and the primed quark fields denote
interaction eigenstates, which differ from the mass eigenstates in
general. The symmetry breaking also leaves a color-octet scalar boson
in the low-energy theory.  The Higgs field $\Phi_{2,3}$ contains two
color-octets under the diagonal SU(3)$_c$; one of them is eaten by the
vector boson $G_H$, while the other remains as a physical scalar multiplet.

The hierarchy of scales in the model is rendered natural through
supersymmetry.  The model can be easily supersymmetrized by introducing
Higgs chiral superfields $\Phi_{2,3}$ transforming as ({\bf 1},{\bf
3},{\bf 3$^*$}) and $\Phi'_{2,3}$ as ({\bf 1},{\bf 3$^*$},{\bf 3}).  The
most general superpotential below the scale $\Lambda_H$ is
\begin{equation}
W = \mu \Tr \Phi \Phi' + h \Det \Phi + h' \Det \Phi',
\end{equation}
with a mass parameter $\mu$, and coupling constants $h$, $h'$.  It is
amusing to note that $\Det\Phi$ is a renormalizable interaction
only for the SU(3)$\times$SU(3) case.  One can easily verify that this
superpotential allows a desired minimum
\begin{equation}
\Phi = v \left( \begin{array}{ccc} 1&&\\&1&\\&&1 \end{array} \right),
\hspace{2cm}
\Phi' = v' \left( \begin{array}{ccc} 1&&\\&1&\\&&1 \end{array} \right),
\end{equation}
where
\begin{equation}
v = h^{-2/3} h^{\prime -1/3} \mu, \hspace{2cm}
v' = h^{-1/3} h^{\prime -2/3} \mu .
\end{equation}
A color-octet supermultiplet is absorbed into the heavy vector
multiplet, leaving two singlet chiral supermultiplets and one color-octet
chiral supermultiplet.  Hereafter, we will only refer explicitly to the
supersymmetric generalization of the model when the superparticle content
has a significant impact on the model's phenomenology.

Given the gauge symmetry of the full theory it is clear that
we can only generate diagonal entries in the quark mass matrices.
In order to generate realistic Cabibbo--Kobayashi--Maskawa (CKM) mixings,
however, we require off-diagonal entries in the basis of the interaction
eigenstates as well.  The way we can generate these off-diagonal components
is through the exchange of heavy vector-like quarks \cite{fn}. We introduce a
minimal set of vector-like quarks, $U$, $C$ and $D$, which have the same
quantum numbers as the ordinary right-handed quarks $u_R$, $c_R$, and
$d_R$.  Without a loss of generality, we can choose a basis where the invariant
Dirac mass terms exist only for the $U$, $C$, and $D$ fields and in which there
is no mass mixing terms between $U$ and $u_R$, etc.  Given the Higgs
representations described earlier, the most general set of Yukawa couplings
involving the vector-like quarks is given by
\begin{eqnarray}
\lefteqn{
{\cal L} = \bar{u}'_R \Phi_{1,2} C + \bar{U} \Phi_{1,2} c'_R
      + \bar{C} \Phi_{2,3} t'_R + \bar{D} \Phi_{1,2} s'_R } \nonumber \\
& & + \bar{Q}_1 H D + \bar{Q}_1 \tilde{H} U
      + \bar{Q}_2 \tilde{H} C + h.c.,
\end{eqnarray}
where $H$ is the Higgs doublet of the minimal Standard Model, and
$\tilde{H} = i \sigma_2 H^*$.  It is straightforward to identify which
$\Phi$ has to be replaced by $\Phi^{\prime *}$ in the supersymmetric
case; $H$ and $\tilde{H}$ must be replaced by $H_1$ and $H_2$ of the
Minimal Supersymmetric Standard Model (MSSM) as well.  After integrating out
the vector-like quarks and replacing the Higgs fields
$\Phi_{1,2}$, $\Phi_{2,3}$ and $H$ by their respective vacuum expectation
values, we obtain the following effective mass terms
\begin{eqnarray}
{\cal L} & = & (\bar{u}'_L, \bar{c}'_L, \bar{t}'_L) M_u
            \left( \begin{array}{c} u'_R \\ c'_R \\ t'_R
                   \end{array} \right)
      + (\bar{d}'_L, \bar{s}'_L, \bar{b}'_L) M_d
                \left( \begin{array}{c} d'_R \\ s'_R \\ b'_R
                       \end{array} \right) + h.c. ,
\end{eqnarray}
where the mass matrices have the form
\begin{equation}
M_u = \left( \begin{array}{ccc}
      * & * & 0 \\ {*} & * & *\\ 0 & 0 & *
           \end{array} \right) , \hspace{1cm}
M_d = \left( \begin{array}{ccc}
      * & * & 0\\ 0 & * & 0\\ 0 & 0 & *
           \end{array} \right).
\end{equation}
Here each asterisk indicates a nonvanishing entry.  By an appropriate
choice of the vector-like quark masses and Yukawa couplings, we
can obtain the following hierarchical form of the quark mass matrices
without any fine-tuning of parameters
\begin{eqnarray}
M_u &\simeq& \left( \begin{array}{ccc}
        m_u & m_c \lambda (\rho - i \eta) & 0\\
      {\cal O} (m_c \lambda^2) & m_c & m_t A \lambda^2\\ 0 & 0 & m_t
             \end{array} \right) \,\, , \\
M_d &\simeq& \left( \begin{array}{ccc}
        m_d & m_s \lambda & 0\\ 0 & m_s & 0\\ 0 & 0 & m_b
             \end{array} \right),
\end{eqnarray}
where we have written our result in terms of the Wolfenstein parameterization
of the CKM matrix.  The (2,1) entry in $M_u$ does not give us a physically
significant effect since it can be eliminated by a rotation on $u_R$
and $c_R$ below the scale $\Lambda_H$.
\footnote{This rotation leads to $D^0$-$\bar{D}^0$ mixing by the exchange
of yet another color-octet vector boson coming from the
breaking $SU(3)_1 \times SU(3)_2 \rightarrow SU(3)_{1,2}$.  However, we
assume that this breaking scale is very high $> {\cal O}(100)$~TeV to evade
constraints from flavor-changing neutral currents processes.}  Note that all
masses and mixing parameters of the minimal Standard Model appear in the
mass matrices above, and it is straightforward to check that they
reproduce the correct form of the CKM matrix.\footnote{Therefore, the form
of the mass matrix is not a {\it predictive}\/ ansatz despite an interesting
texture.} In fact, we can even drop the (2,1) entry completely,
and still reproduce the correct CKM matrix.

One of our primary interests, however, is to consider the effects of our
model on $B$-physics.  Even though the mass matrices that we have constructed
can perfectly reproduce the CKM matrix, they do not at the moment have
any effect on $B$-physics because the first two generations are decoupled
from the third in $M_d$; when we go to the mass eigenstate basis, we do not
generate any new $\Delta B=1$ operators via $G_H$ exchange.  Therefore, we
introduce an additional vector-like quark $B$ that has the same quantum
numbers as $b_R$.  The new Yukawa couplings involving the $B$ field
are given by
\begin{equation}
\Delta {\cal L} = \bar{s}'_R \Phi_{2,3} B + \bar{Q}_3 H B .
\end{equation}
After integrating out the $B$ field, we obtain a single new entry in the
down-quark mass matrix, which we assume is of the order $m_b \lambda$ or
less. Thus,
\begin{equation}
M_d \simeq \left( \begin{array}{ccc}
        m_d & m_s \lambda & 0\\ 0 & m_s & 0\\ 0 & m_b \lambda \xi & m_b
             \end{array} \right),
\end{equation}
where $|\xi| \lesssim {\cal O}(1)$ is a new complex parameter.

It is important to note that the mass matrices we have constructed not
only reproduce the correct CKM matrix (despite their many zeros) but they
also lead to relatively small FCNC effects from $G_H$ exchange,
as we will see in the next two sections.  In addition, our form of the mass
matrices have the nice feature that Cabbibo mixing is generated in the
down-sector while $V_{cb}$ is generated in the up-sector.  This is
preferred from the point of view of model building, since it leaves open
the possibility of explaining the famous relation $\lambda \simeq
\sqrt{m_d/m_s}$, and perhaps also $V_{cb} \simeq \sqrt{m_c/m_t}$.  It is
an interesting direction for future research to see whether it is possible
to construct a realistic and predictive model within this framework,
though we do not pursue this issue further here.

\section{Parameter Space} \label{sec:param}

In Figure~1 we show the allowed region of the $M$-$\cot\theta$ plane. The
s-channel $G_H$ exchange diagram can contribute significantly to the
$t\overline{t}$ production cross section.  We have adopted the central value of
the $t\overline{t}$ cross section expected in the Standard Model from the
next-to-leading order calculation by Laenen {\em et al.}\cite{laenen}, and have
computed the shift of the production cross section in the $q\overline{q}$
channel expected in our model, including interference effects. The partonic
cross section in the $q\overline{q}$ channel is given by
\beq
\hat{\sigma}(q\overline{q}\rightarrow t\overline{t})=
\frac{8\pi}{27} \alpha_s^2 \frac{1}{\hat{s}} \sqrt{1-\frac{4 m_t^2}
{\hat{s}}} \left[1+\frac{2 m_t^2}{\hat{s}}\right] R
\eeq
where $\hat{s}$ is the parton center of mass energy squared, and
$R$ is given by
\beq
R \equiv \frac{M^2(M^2+\Gamma^2)}{(\hat{s}-M^2)^2+M^2 \Gamma^2}
\eeq
The $G_H$ width $\Gamma$ is given by $\alpha_s M (4 \tan^2\theta+
\cot^2\theta)/6$, for decay to $5$ light flavors, for example.
We use the MRS D- structure functions in computing the total
$q\overline{q}$ cross section, to be consistent with
Ref~\cite{laenen} , and have assumed a $K$-factor of 1.2 to take into
account next-to-leading-order effects. \footnote{In fact, Laenen {\em et al.}
use MRS D-' structure functions, but the difference is negligible for
$t\bar{t}$ production.}   We require the total cross section to be within
the 95\% confidence level bounds of the value observed by
CDF, $\sigma (t\bar{t}) = 6.8^{+3.6}_{-2.4}$ pb, for $m_t=176 \pm 13$
GeV\cite{cdf}.  Because of the small statistics, the experimental errors
are not exactly gaussian and there is a correlation between the errors
in $\sigma (t\bar{t})$ and $m_t$.  Unfortunately, the CDF paper \cite{cdf}
does not show the variation of $\chi^2$ over an extended range of $m_t$,
nor the dependence of the efficiency on $m_t$. Thus, we have no way
to determine the 95\% confidence level bound rigorously. We simply double
the error bar on the CDF cross section to obtain an allowed range,
$2$--$14$ pb, for $m_t$ between $150$--$202$~GeV.  Also, we do not take the
correlation of the errors into account, and vary both $\sigma (t\bar{t})$
and $m_t$ within the above ranges independently.\footnote{Even though this
``rectangular'' treatment of the errors correspond to $(95\%)^2 =
90~\%$ confidence level, the final bound is determined by the corners of
the rectangle, and hence correspond to 98~\% confidence level in the
(uncorrelated) Gaussian case.  The true confidence level depends on both
the correlation and non-Gaussian nature of the errors.  Therefore, the
excluded region in our figure should only be thought of as an approximate
bound.}  We assumed a theoretical uncertainty in the central value of the
Standard Model $t\overline{t}$ cross section that follows from varying the
renormalization scale between $m_t/2$ and $2 m_t$.  The excluded region
lies within the oval area between $\sim$400 and $\sim$800 GeV.

The dashed line in Figure~1 shows the bounds on new particles decaying
to dijets at the 95\% confidence level, from both UA1 \cite{ua1}
(between 200 and 300 GeV) and CDF \cite{cdf2} (between 200 and 850 GeV).
The $G_H$ decays to quark-antiquark pairs, like the more familiar axigluon,
but its coupling to light quarks is suppressed by a relative factor
of $1/\cot\theta$.  Thus, given the published bounds on the axigluon
production cross section, we can determine the value of $\cot\theta$
necessary to suppress the $G_H$ cross section until it is below the
experimental bound.  From this, we conclude that the region below the
dashed line is excluded.  Note that the older UA1 bounds are more
stringent below $\sim$250 GeV due to higher statistics at smaller values
of $\hat{s}$.

We also show the region allowed by the $Z\rightarrow b\overline{b}$
width measured at LEP.  The new contribution to the parameter
$R_b\equiv \Gamma(Z \rightarrow b\overline{b}) /
\Gamma(Z \rightarrow \mbox{hadrons})$ is given by
\beq
\frac{\Delta R_b}{R_b}= \frac{\frac{2}{3\pi} \alpha_s
(\cot^2\theta-\tan^2\theta) (1-R_b) F(M,M_Z)}{
1+\frac{2}{3\pi} \alpha_s
\left[\tan^2\theta + (\cot^2\theta-\tan^2\theta) R_b\right] F(M,M_Z)}
\label{eq:rb}
\eeq
where the function $F$ is provided in Appendix~A.  Note that there
is another contribution in the supersymmetric version of the model which
involves the superpartners of the $G_H$-boson and the bottom
quark.  However, this diagram is roughly one order of magnitude
smaller than the result in (\ref{eq:rb}) \cite{amp}.\footnote{Of course
there could be an effect from the ordinary particle content
of the MSSM, but it is negligible unless both the top squark and the
chargino masses are just beyond the LEP limits.} Again we require
that the Standard Model value of $R_b$ plus the contribution given above
remain within the 95\% confidence level bound of the value measured at LEP,
0.2210$\pm$0.0029 \cite{pdg}.  The dotdashed line indicates the experimental
central value, while the solid lines put an upper and lower bound
on $\cot \theta$.  Since there is a small discrepancy between the
measured value from the Standard Model prediction, the central value lies
away from the line $\cot \theta = 1$ where the effect vanishes.
\footnote{In evaluating (\ref{eq:rb}), we have taken $\alpha_s(m_Z)=0.110$,
as determined by lattice QCD, rather than the LEP value.  This is because
the observed enhancement in $Z\rightarrow b\overline{b}$ is roughly in a one
to one correspondence with the high value of $\alpha_s(m_Z)$ observed at
LEP.   However, it does not change our figure significantly if we
choose $\alpha_s (m_Z) = 0.120$ instead.  If we had used the preliminary
number $R_b = 0.2204 \pm 0.0020$ presented at Moriond, which is 2.4~$\sigma$
off from the Standard Model prediction, the allowed range lies in
a narrower band above $\cot \theta = 1$ everywhere, while the curve
for the central value does not shift significantly.}  Note that the
analogous quantity defined for charm quarks, $R_c$, is reduced in this model
if $\cot \theta > 1$:
\beq
\frac{\Delta R_c}{R_c}
= - \frac{R_b}{1-R_b} \frac{\Delta R_b}{R_b} .
\eeq
Thus, when we account for the central value of $R_b$ observed
at LEP, we also obtain an $-$0.7\% shift in $R_c$, which is
completely negligible compared to the experimental error, $R_c =0.171 \pm
0.020$ \cite{pdg}.  Other $Z$-pole observables do not give us any
further constraints.   For instance, the forward-backward and
polarization asymmetries in $Z \rightarrow b \overline{b}$  are
unaffected in our model because the $G_H$ couples equally to left- and
right-handed quarks.

Finally, we show the region excluded by flavor changing neutral current
bounds in the case where the parameter $\xi=0$.  The case of
nonvanishing $\xi$ will be discussed together with the associated $B$ physics
in the next section.  Given the minimal ($\xi=0$) form of the mass matrices
described in Section 2, there is no constraint from the down-sector, and
the largest flavor changing neutral current bound comes from the
$D^0$-$\overline{D}^0$ mass splitting.  The effect is given
approximately by
\beq
\Delta m_D = \frac{2 \pi}{9} \alpha_s
\frac{\lambda^{10} A^4 (\rho^2 + \eta^2)}
      {M^2 \sin^2 \theta \cos^2 \theta}
      B_D f_D^2 m_D \eta_{QCD}
\label{eq:ddbar}
\eeq
where $f_D = 208 \pm 37$ is the $D^0$ decay constant as determined
from a lattice QCD calculation \cite{soni}, and
$\eta_{QCD} \approx 0.58$ is the QCD correction from running the
four-quark operator down from $M_Z$ to $m_D$, using the conventions of
Buras {\em et al.} \cite{Buras}.  \footnote{We consistently ignore the
effects of running between $\mu = M$ and $m_Z$, since
$\log(m_Z/m_b) \gg \log(M/m_Z)$ and
$\alpha_s(\mu>m_Z)< \alpha_s(\mu<m_Z)$ as well.}
We take $B_D(m_c) = 1$ in our numerical analysis. We
require that (\ref{eq:ddbar}) is less than the experimental
upper bound $\Delta m_D< 1.32 \times 10^{-10}$ MeV to determine the
excluded region of the parameter space.  This is appropriate because the
Standard Model contribution to $D^0$-$\overline{D^0}$ mixing is expected
to be $\sim 5$ orders of magnitude smaller than the current experimental
limit \cite{gsor}.  The excluded region lies to the left of the parabolic
solid line shown in Figure~1.

The exchange of the color-octet Higgs boson also contributes to
flavor-changing processes, but with a much smaller effect.
The $\bar{t}'_R c'_L \Phi_{2,3}$ coupling is of order $m_t A \lambda^2 /
\langle \Phi_{2,3} \rangle$, and the coupling to the mass eigenstates
$\bar{c}_R u_L$ is suppressed by an additional factor of
$A\lambda^2 (m_c/m_t) \times \lambda (\rho - i\eta)$.  Thus, the
contribution to $\Delta m_D$ from Higgs exchange is down by at least
$(m_c/m_t)^2$ compared to (\ref{eq:ddbar}).  We also checked that the
exotic decay modes $t\rightarrow c g$, $c\gamma$ and $c\bar{b} b$ are
negligible compared to the Standard Model decay $t \rightarrow b W$.

Taking all the constraints into account, we see from Figure~1 that there
are two allowed ``windows'', in our model's parameter space: one small
region at approximately $M\approx 250$ GeV, and a somewhat larger region
above $M\approx 600$ GeV.  If we further require that the central
experimental value of $R_b$ from LEP has to be reproduced, we must take
$\cot \theta \sim 2$ in the light window and $\cot \theta \sim 4$
in the heavy window.

\section{$B$-physics} \label{sec:bphys}

Using the $\xi=0$ form of the down-quark mass matrix, we showed in
the previous section that we could account for $R_b$ within the
two allowed regions of our model's parameter space.  In this
section, we will consider the effects of the nonminimal ($\xi \neq 0$)
form of $M_d$ on $B$-physics.  The question we would like to address is
whether we can also reduce the $B$-meson semileptonic branching
ratio $B_{SL}$ and the charm multiplicity per decay $n_c$ through
the additional $\Delta B =1$ operators that are present
when $\xi \neq 0$, without conflicting with the constraints from
FCNC processes.  We will argue that our results strongly favor
the supersymmetric version of the model.

With $\xi \neq 0$, we generate the following effective $\Delta B = 1$
operator through the exchange of the $G_H$ boson.
\begin{equation}
{\cal L}_{\it eff} = \frac{g_s^2}{M^2} \frac{\lambda\xi}{\cos^2 \theta}
      \bar{b}_R \gamma^\mu T^a s_R
      ( \bar{u} \gamma^\mu T^a u + \bar{d} \gamma^\mu T^a d
            + \bar{s} \gamma^\mu T^a s + \bar{c} \gamma^\mu T^a c) ,
            \label{bsqq}
\end{equation}
Since this operator always has an $s_R$ in the final state, the amplitude
does not interfere with the Standard Model amplitude (which
only involves $s_L$) providing that we neglect $m_s$ compared to $m_b$.
Therefore, we add the new decay amplitude incoherently to that of the
Standard Model.  The QCD running modifies the normalization of
the operator from the weak scale to the scale $m_b$; we take this
into account in our numerical analysis below.

We estimate the additional contributions to the $B$ partial decay widths
from the effective operator in (\ref{bsqq}) using the parton-level spectator
quark approximation.  We find
\begin{equation}
\Delta \Gamma_B = \frac{m_b^5}{192 \pi^3} \frac{N_C^2-1}{16N_C}
      \left(
            \frac{g_s^2 \lambda \xi}{M^2 \cos^2 \theta} \right)^2
      [ 3 G(0) + G(4 m_c^2/m_b^2) ] ,
\end{equation}
where the function $G$ is defined by
\begin{equation}
G(y) = \frac{1}{16} \sqrt{1-y} (16 - 40y + 18 y^2 - 9y^3)
      + \frac{3}{32} (8 - 3y) y^3
            \ln \frac{1 + \sqrt{1-y}}{1 - \sqrt{1-y}} ,
\end{equation}
and is normalized such that $G(0) = 1$.

As we described earlier, the current experimental data suggests that
both the $B$ semileptonic branching ratio $B_{SL}$, and the
charm multiplicity per $B$ decay $n_c$, are lower than the Standard
Model expectations.  Notice that decays via the operator (\ref{bsqq})
increase the total decay rate without affecting the semileptonic mode,
and hence $B_{SL}$ decreases.  This operator also reduces $n_c$ since
it contributes to the $sq\bar{q}$ final state, which involves
charm quarks only part of the time.  We list the effects on $B_{SL}$ and $n_c$
for various choices of the parameters in Table.~1.  The effect of $G_H$
exchange is described by the parameter $\Delta$, defined by
\begin{equation}
\Delta = \left( \frac{750~\mbox{GeV}}{M} \right)^4
      \frac{\xi^2}{\cos^4 \theta} .
\end{equation}
We used the Standard Model predictions quoted in Ref.~\cite{Altarelli}
with the renormalization scale $\mu = m_b$.  For $\mu =
m_b/2$, $B_{SL}$ decrease by only 4\%, while $n_c$ does not change at
all.  The ``heavy'' case corresponds to $m_c = 1.7$~GeV, $m_b =
5.0$~GeV, and the ``light'' case to $m_c = 1.2$~GeV, $m_b =
4.6$~GeV.  QCD renormalization effects were taken into account by
numerically solving the renormalization group equation using the
anomalous dimension matrix given in Ref.~\cite{GSW}.  For the ``heavy ''
quark masses it is difficult to reproduce the experimental values for any
choice of $\Delta$.  However, one must keep in mind that the value
of $B_{SL}$ corresponding to the ``heavy'' quark masses is very far from the
experimental data for $\Delta=0$, indicating a more serious problem
from the beginning.  For ``light'' quark masses the discrepancy with the
Standard Model predictions are $\sim 20$\% for both $B_{SL}$ and $n_c$,
and we can easily obtain these shifts by an appropriate choice of the mixing
parameter $\xi$, providing we do not exceed the bounds on other flavor
changing processes, as we describe below.
\begin{table}
\centerline{
\begin{tabular}{l|cc|c}
masses&\multicolumn{2}{c|}{heavy}&experiment\\
\hline
$\alpha_s (m_Z)$ & 0.105 & 0.115 &\\ \hline
$B_{SL}$ & $15.4 (1-0.22 \Delta)$ & $12.7 (1-0.28 \Delta)$
      & $10.6\pm 0.3$ \cite{Altarelli}\\
$n_c$ & $1.09 (1 - 0.21 \Delta)$ & $1.09 (1 - 0.25 \Delta)$
      & $1.08 \pm .06$ \cite{muheim}
\end{tabular}
}
\caption{Effect of $G_H$ exchange on $B$-decay for the choice of
``heavy'' masses, $m_c = 1.7$~GeV and $m_b = 5.0$~GeV.  The
renormalization scale of the Standard Model operator is chosen at $\mu =
m_b$.}
\end{table}

\begin{table}
\centerline{
\begin{tabular}{l|cc|cc|c}
masses&\multicolumn{2}{c|}{light}&experiment\\
\hline
$\alpha_s (m_Z)$ & 0.105 & 0.115 &\\ \hline
$B_{SL}$ & $13.2 (1 - 0.16 \Delta)$ & $12.7 (1 - 0.20 \Delta)$
      & $10.6\pm 0.3$ \cite{Altarelli}\\
$n_c$ &     $1.21 (1 - 0.13 \Delta)$ & $1.21 (1 - 0.16 \Delta)$
      & $1.08 \pm .06$ \cite{muheim}
\end{tabular}
}
\caption{Effect of $G_H$ exchange on $B$-decay for the choice of
``light'' masses, $m_c = 1.2$~GeV and $m_b = 4.6$~GeV.  The
renormalization scale of the Standard Model operator is chosen at $\mu =
m_b$.}
\end{table}

Now that we have introduced mixing between $b_R$ and $s_R$, we must
also consider the constraints from the other FCNC processes
associated with a nonvanishing value of $\xi$.  We found that the most
stringent constraint comes from the contribution
to $B^0$-$\overline{B}^0$ mixing from the operator
\begin{equation}
{\cal L} = \frac{1}{2} \left( \frac{g_s}{M \sin \theta \cos \theta}
      U_{31}^{d_R} \right)^2
      \bar{b}_R \gamma^\mu T^a d_R \bar{b}_R \gamma_\mu T^a d_R ,
\end{equation}
Here, $U^{d_R}$ is the rotation matrix acting on the right-handed down-type
quarks, and $U_{31}^{d_R} \simeq \lambda \xi (\lambda m_d/m_s) \simeq
\lambda^4 \xi$ is the mixing angle between $b_R$ and $d_R$. The
renormalization of this operator down to the scale $m_b$ gives us the
correction factor $\eta_{QCD} \simeq 0.55$, again using the definition of
$\eta_{QCD}$ given in Ref~\cite{Buras}. The contribution to the mass
splitting is then given by
\begin{equation}
\left. \Delta m_B \right|_{G_H} = \frac{2 \pi}{9} \alpha_s
      \left( \frac{\lambda^4 \xi}{M \sin \theta \cos \theta}
      \right)^2
      f_B^2 B m_B^2 \eta_{QCD} \ ,
\end{equation}
where $B f_B^2  = (1.0 \pm 0.2) (180 \pm 50~\mbox{MeV})^2 $ ,
based on recent lattice estimates \cite{glas}. As before, we
obtain much smaller effects from octet Higgs exchange.  The sum of
the Standard Model contribution $\left. \Delta m_B
\right|_{SM}$ and $\left. \Delta m_B \right|_{G_H}$ shown above should
give us the experimental value
\beq
\left| \left. \Delta m_B \right|_{SM} +
      \left. \Delta m_B \right|_{G_H}
\right|
    = \left. \Delta m_B \right|_{expt}
\eeq
where $\left.\Delta m_B \right|_{expt} = 3.357\times 10^{-13}$~GeV.
However, the  Standard Model prediction depends sensitively on the
choice of $V_{td}$.  If we fix $m_t = 176$~GeV and
$B f_B^2 = (180~\mbox{MeV})^2$, then $\left. \Delta m_B \right|_{SM}$
can range between (0.16--2.50) $\left. \Delta m_B \right|_{expt}$ \cite{glas}
if we allow $V_{td}$ to vary within the 90\% C.L. range allowed by
CKM unitarity $V_{td} = 0.004$--$0.015$ \cite{pdg}.  Since the phase of the
parameter $\xi$ is arbitrary, we obtain the bound
$\left| \left. \Delta m_B \right|_{G_H} \right| < 3.5 \left. \Delta m_B
\right|_{expt}$, and hence
\beq
M< \frac{3.11 \mbox{ TeV}}{\sqrt{\Delta}} \frac{1}{\cot\theta} \ .
\label{eq:bnd}
\eeq
Since we have used the 90\% C.L. upper bound on $V_{td}$ in deriving this
expression while keeping $m_t$ and $B f_B^2$ fixed at their central values,
it follows that the actual 95\% C.L. bound is weaker than the one given
in (\ref{eq:bnd}).  We use this strict bound to demonstrate that
even with conservative assumptions there exists an allowed range
in the parameter space of the model when $\Delta \approx 1$.  If we
choose $\Delta = 1$, so that we can account for a 20\% reduction
in $B_{SL}$ and $n_c$, then we must also be below the dashed line shown
in Figure~2 to evade the $B$-$\overline{B}$ mixing constraint given
by (\ref{eq:bnd}).  Thus far, we see that both in the light and heavy
windows, we can explain the anomalies in $B_{SL}$, $n_c$, and $R_b$
simultaneously.

The only other potentially significant FCNC constraint that depends on the
parameter $\xi$ and that may alter our conclusions is the $b\rightarrow s
\gamma$ branching fraction.  Note that {\em all} of the flavor-changing
processes that we have considered until now have received contributions at
tree-level in our model, while the $b\rightarrow s \gamma$ width receives
contributions at the one-loop level only.  This point is significant because
the supersymmetric particle content of our model can now have a dramatic effect
on the result.

Let us consider the upper bound that we can place on
$\cot\theta$ from the $b\rightarrow s \gamma$ branching fraction.  The  most
important contributions to the amplitude for $\cot\theta > 1$ come from the
Feynman diagrams proportional to $1/\sin^2\theta$, shown in Figure~3.  If we
first consider the non-supersymmetric case, the entire nonstandard contribution
to the inclusive branching fraction comes from the first diagram.   If we
compare this to the difference between the CLEO upper bound,
$4.2 \times 10^{-4}$ \cite{CLEO} and the Standard Model expectation,
$1.9 \pm 0.54$ \cite{ciu}, and require agreement within two
standard deviations, then we obtain the constraint
\beq
\cot\theta<1.7/\sqrt[4]{\Delta}
\label{eq:nsb}
\eeq
If we require $\Delta=1$ to explain the $B$-decay anomalies, then
we see that (\ref{eq:nsb}) forces us to live in the lower right
handed corner of the allowed region in Figure~2, where we cannot
simultaneously explain the central value of $R_b$.

However, the situation changes dramatically in the supersymmetric case.  It is
well known that the $b\rightarrow s \gamma$ branching fraction vanishes
identically in the limit of exact supersymmetry \cite{susy}.  Thus, the new
contribution to $b\rightarrow s \gamma$ in our model not only depends on $M$
and $\cot\theta$, but also on the soft-SUSY breaking masses and couplings.  In
addition to the diagram involving ordinary particles, there are a number of
other diagrams that involve the exchange of the left- and right-handed
$b$-squarks, ($\tilde{b}_L$, $\tilde{b}_R$) and the fermionic partners of the
transverse and longitudinal components of the $G_H$-boson ($\psi$, $\chi$).
The new diagrams tend to cancel the first one.  An exact result for these
diagrams is provided in Appendix~B.  If we assume that the soft SUSY-breaking
masses defined in Appendix~B are around $100$ GeV, temporarily ignore
$\tilde{b}_L$-$\tilde{b}_R$ mixing effects (i.e. ignore the fourth diagram),
and compare the total result for the inclusive branching fraction to the CLEO
upper bound, we now find
\begin{eqnarray}
\cot\theta&<&3.2/\sqrt[4]{\Delta} \,\,\,\,\,\,\ (M=250 \mbox{ GeV}),
\nonumber \\
\cot\theta&<&5.5/\sqrt[4]{\Delta} \,\,\,\,\,\,\ (M=640 \mbox{ GeV}).
\label{eq:set1}
\end{eqnarray}
correspoding to bound (a) in Figure~2. In this case we do not exclude the
interesting regions of our parameter space when $\Delta=1$.  We show the
regions excluded for $\Delta=1$ and for various choices of the SUSY breaking
parameters in Figure~2. Note that we have not included the QCD running effects
in these bounds since the relevant anomalous dimension matrix is not available
in the literature.  While this effect may give an important correction to
our estimates above, our conclusions will not change once we take into account
the strong dependence of our results on the soft supersymmetry-breaking
parameters.  For example the running of the $b\rightarrow s\gamma$ operator
itself results in a suppression in the amplitude by a factor of $\sim 0.7$,
while the four-fermi operator Eq.~(\ref{bsqq}) may feed into the $b\rightarrow
s\gamma$ operator at the two-loop level and enhance the amplitude by a factor
of 2 or 3. However, if we include the $\tilde{b}_L$-$\tilde{b}_R$ mixing in the
fourth diagram, it is easy to obtain a suppression of approximately the same
size without any real fine-tuning ({\em e.g.} the difference between bounds (b)
and (c) in Figure~2).  This is possible because we do not know the magnitude or
phase of the soft SUSY-breaking parameter $A$.  With $A\approx 100$ GeV, the
bounds in (\ref{eq:set1}) become
\begin{eqnarray}
\cot\theta&<&3.8/\sqrt[4]{\Delta} \,\,\,\,\,\,\ (M=250 \mbox{ GeV}),
\nonumber \\
\cot\theta&<&8.6/\sqrt[4]{\Delta} \,\,\,\,\,\,\ (M=640 \mbox{ GeV}).
\end{eqnarray}
While a complete study of this enlarged parameter space
is beyond the scope of this paper, it is clear that the suppression
of $b \rightarrow s \gamma$ in the supersymmetric case, as well as
the additional parametric degrees of freedom prevent us from
excluding additional regions of the $M$-$\cot\theta$ plane
when $\Delta=1$ ({\em c.f.} bound (d) in Figure~2). Thus, we can
account for the observed value of $R_b$, and obtain a 20\% reduction
in $B_{SL}$ and $n_c$ simultaneously.

\setcounter{footnote}{0}

\section{Confinement Phase of SU(3)$_3$} \label{sec:conf}

We have seen that it is possible to explain the anomalies in $R_b$,
$B_{SL}$ and $n_c$ in both allowed regions of our model's parameter
space.  However, the strong possibility that the lighter window
($M \approx 250$ GeV) may be ruled out in the near future by the
improving bounds on new particles decaying to dijets makes the heavier
region ($M \gtrsim 600$~GeV) worthy of further consideration. In order
to reproduce the central value of $R_b$ in this region would require that
we take $\cot \theta \sim 4$, which implies a rather large $SU(3)_3$ gauge
coupling.  This leaves open the possibility of nontrivial nonperturbative
dynamics.  For example, if the dynamics of the third family in the limit
of large $g_3$ is like the Nambu--Jona-Lasinio model, then we would
expect chiral symmetry breaking at a critical coupling of
$\cot\theta = \sqrt{\pi/(3\alpha_s)} \approx 3.4$, and a large dynamical
mass for both the top and bottom quarks. This would not be
phenomenologically acceptable.  However, what actually happens in
this limit depends on the nonperturbative dynamics and on the mass of
the Higgs boson $\Phi_{2,3}$.  In this section, we consider the
possibility that the SU(3)$_3$ gauge theory may be confining, and that
the physical bottom and top quarks may be composite particles, bound
states of the fundamental quark fields and the Higgs field
$\Phi_{2,3}$.  We will argue that the phenomenology of such a confining
phase is the same as the Higgs phase we have described in the
previous three sections, providing that the third family dynamics is
similar to the Abbott--Farhi model~\cite{afar}.

Let us consider what happens if $g_3 \gg 1$ and SU(3)$_3$ is confining. If
$\Phi_{2,3}$ is much heavier than the scale at which $g_3$ becomes large, then
$\Phi_{2,3}$ decouples from the dynamics and we expect spontaneous chiral
symmetry breakdown, and a large dynamical mass for the third generation quarks.
If $\Phi_{2,3}$ is lighter than the scale at which the $SU(3)_3$ coupling
becomes strong, then complementarity \cite{compl} suggests that the resulting
confinement phase should be smoothly connected to the Higgs phase described
earlier in this paper.  Thus, we expect the physical top and bottom quarks to
become composites of the fundamental top and bottom quarks with the
$\Phi_{2,3}$ boson.  We identify physical composite top and bottom states,
$t_c$ and $b_c$, with the composite operators $(\Phi t)/\Lambda$ and $(\Phi
b)/\Lambda$, where $\Lambda$ is the compositeness scale. Note that these
composite states transform as triplets under $SU(3)_{1,2}$ as desired.   In
this phase it is not necessary to have $\langle \bar{t}t \rangle$ or $\langle
\bar{b} b \rangle$ condensates since the low-energy fermion content satisfies
the 'tHooft anomaly matching conditions, as in the Abbott--Farhi model
\cite{afar}. There is a $\rho$-like meson state in the confining phase composed
of $t\bar{t}$, $b\bar{b}$ and $\Phi \Phi^*$ bound states, which corresponds to
the $G_H$ boson of the Higgs phase.  The composite top and bottom quarks would
remain massless if it were not for the explicit chiral symmetry breaking
effects of their associated Yukawa couplings, and hence the masses of the
composite states are proportional to their Yukawa couplings as before.

All the phenomenological constraints that we have discussed also carry
over smoothly to the confining picture.   For example, let us consider
the $q\overline{q} \rightarrow t\overline{t}$ cross
section presented in Section~\ref{sec:param}.  The only contribution
to this process in the confining picture is s-channel gluon exchange,
but now there is a form factor $F(q^2)$ at the gluon-$t_c\overline{t}_c$
vertex.  Assuming vector-meson dominance, $F(q^2)$ is saturated by the
exchange of the $\rho$-like meson ($G_H$), which mixes with the gluon.
Since $F(q^2)$ should reduce to the form factor for a point-like
interaction $F(q^2) \rightarrow 1$ in the $q^2 \rightarrow 0$ limit, it
must be of the form
\begin{equation}
F(q^2)=\frac{-M^2 + i M \Gamma}{q^2 - M^2 + i M \Gamma}
\end{equation}
for small $q^2$.  Thus, the cross section computed in the confining phase has
exactly the same dependence on $M$ and $\Gamma$ as the cross section we
obtained from perturbative $G_H$ exchange in the Higgs phase.  We can also
describe the phenomenology of the confining phase in terms of the same
parameter space as the Higgs phase, providing that $q^2 \ll M^2$. The
$\rho$-like meson does not couple to the light quarks directly, while the
coupling to $t_c$ and $b_c$ is an unknown parameter that we can identify with
$g_s \cot\theta$. Therefore, the correction to $R_b$ involves the same
combination of parameters as before.  The dimension-five operators generated by
the exchange of vector-like fermions induce mixing between elementary light
quarks and composite third generation quarks, and the flavor off-diagonal
coupling of the $\rho$-like meson is proportional to $g_s \cot\theta$, again
consistent with the Higgs phase result in the limit $\cot \theta \rightarrow
\infty$.

\section{Conclusions} \label{sec:conc}

We have shown that the extension of the Standard Model presented in this paper
can yield corrections to $R_b$, the $B$-meson semileptonic branching ratio and
the charm multiplicity in $B$-decays, in a pattern roughly consistent with the
experimental data.  In the nonsupersymmetric version of the model, we can
explain either the value of $R_b$, or the $B$-decay anomalies within the
allowed parameter space of our model, but not both simultaneously.  This is a
consequence of the strong upper bound on $\cot\theta$ that we obtain from the
limit on $b\rightarrow s \gamma$, when we take $\Delta =1$.  However, the
natural suppression of $b \rightarrow s \gamma$ in the supersymmetric version
of the model allows for choices of the parameters that can account for $R_b$
and the $B$-decay anomalies simultaneously.  Because the massive color octet
bosons in this model couple more strongly to third generation quarks, a natural
way to test this idea is to search for a high mass peak in the $b\overline{b}$
dijet invariant mass distribution at hadron colliders.  This may rule out
the small allowed window that we found near $M \approx 250$ GeV in the near
future.\footnote{A search is currently under way at CDF, but the data
analysis has not yet been  completed.}  At present it is interesting to note
that the most recent search at CDF \cite{cdf2} for new particles decaying to
dijets has found upward statistical fluctuations in the data near 250, 550, and
850 GeV \cite{cdf2}.  While probably nothing more than coincidental, it is
amusing to note that two of these masses fall within the two allowed regions in
our model's parameter space.  Finally, we have considered the limit in which
the SU(3) associated with the third family quarks becomes confining, and argued
that the bottom and top quarks may become composite objects without forming
condensates, like the composite fermions of the Abbott-Farhi Model.  We argued
that the phenomenology of the confinement phase should be smoothly connected to
that of the Higgs phase described in this paper.  Thus, it is possible that
third family compositeness effects may give us other nontrivial signatures of
this model.

\begin{center}
{\bf Acknowledgments}
\end{center}

We thank JoAnne Hewett and Tom Rizzo for useful comments.
{\em This work was supported by the Director, Office of Energy Research,
Office of High Energy and Nuclear Physics, Division of High Energy
Physics of the U.S. Department of Energy under Contract DE-AC03-76SF00098.}

\appendix
\section{Appendix}
The function $F(M,M_Z)$ defined in Eq.~\ref{eq:rb} is given by
\beq
F(M,M_Z)\equiv F_1 + F_2
\eeq
for $M<m_Z$ and
\beq
F(M,M_Z)\equiv F_2
\eeq
for $M>m_Z$, where
\begin{eqnarray}
F_{1}&=&(1+\delta)^2 \left[3\ln\delta + (\ln\delta)^2\right]
+5(1-\delta^2)-2\delta\ln\delta
\nonumber \\
& &-2(1+\delta)^2
\left[\ln(1+\delta)\ln\delta+\mbox{Li}_2\left(\frac{1}{1+\delta}\right)-
\mbox{Li}_2\left(\frac{\delta}{1+\delta}\right)\right] , \\
F_{2}&=&-2\left\{\frac{7}{4}+\delta+(\delta+\frac{3}{2})\ln\delta \right.
\nonumber \\
& &\left. + (1+\delta)^2\left[\mbox{Li}_2\left(\frac{\delta}{1+\delta}\right)
+\frac{1}{2}\ln^2\left(\frac{\delta}{1+\delta}\right)
-\frac{\pi^2}{6}\right]\right\} .
\end{eqnarray}
Here Li$_2(x) = -\int_0^x \frac{dt}{t} \ln(1-t)$ is the Spence function,
and $\delta=M^2/m_Z^2$.

\section{Appendix}
The effective operator yielding the largest contribution to
to $b\rightarrow s \gamma$ for $\cot\theta>1$ is given by
\beq
\frac{1}{36\pi^2} \frac{e g_s^2 \lambda \xi}{\sin^2\theta}\frac{m_b}{M^2}
\sum_i \frac{c_i}{2} \, \overline{s}_R \, \sigma^{\alpha\beta} F_{\alpha\beta}
\, b_L
\eeq
where $F_{\alpha\beta}$ is the photon field strength tensor, and
the index $i$ runs over the four classes of diagrams shown in
Figure~3. We find to lowest order in the bottom quark mass:
\begin{eqnarray}
c_1 &=& \frac{4}{3}
\\
c_2 &=&-\sum_i \frac{U_{\psi i} U_{\chi i} M }{\mu_i}
\left[\frac{1-r_i^4+2r_i^2 \ln(r_i^2)}{(1-r_i^2)^3} \right]
\\
c_3 &=& - \sum_i \frac{U_{\psi i} U^\dagger _{\psi i} M^2}{\mu_i^2}
\left[
\frac{\frac{1}{6} p_i^6-p_i^4+\frac{1}{2} p_i^2 + p_i^2\ln(p_i^2)
+\frac{1}{3}} {(1-p_i^2)^4} \right]
\\
c_4 &=& 2 A \sum_i \frac{U_{\psi i} U_{\psi i} M^2}{\mu_i^3}
\left[ \frac{ -4 + (1+p_i^2)(1+r_i^2)}{2(1-p_i^2)^2(1-r_i^2)^2} \right.
\nonumber \\
&&\left. -\frac{r_i^2}{(1-r_i^2)^3(r_i^2-p_i^2)}\ln r_i^2 -
\frac{p_i^2}{(1-p_i^2)^3(p_i^2-r_i^2)}\ln p_i^2 \right]
\end{eqnarray}
where we have neglected terms of order $m_b^2/M^2$. Above the $\mu_i$ are the
eigenvalues of the $\psi$-$\chi$ mass matrix
\beq
\left(\begin{array}{cc}  m_\psi & M \\ M & m_\chi
\end{array}\right)
\label{eq:mmm}
\eeq
where $\psi$ is the superpartner of the the transverse
component of the $G_H$, and $\chi$ is the superpartner
of the would-be Nambu Goldstone boson in $\Phi_{2,3}$.
Note that the off-diagonal entries are generated by the
vacuum expectation value of $\Phi_{2,3}$, while the diagonal
ones are SUSY breaking effects.  $U$ is the matrix that diagonalizes
(\ref{eq:mmm}), $r_i=m_{\tilde{b}_L}/\mu_i$, and $p_i=m_{\tilde{b}_R}/\mu_i$.
In the final diagram, $A m_b$ parameterizes the $\tilde{b}_L$-$\tilde{b}_R$
mass squared mixing, where $A$ is the trilinear soft SUSY breaking parameter.


\newpage
\begin{center}
{\bf Figure Captions}
\end{center}

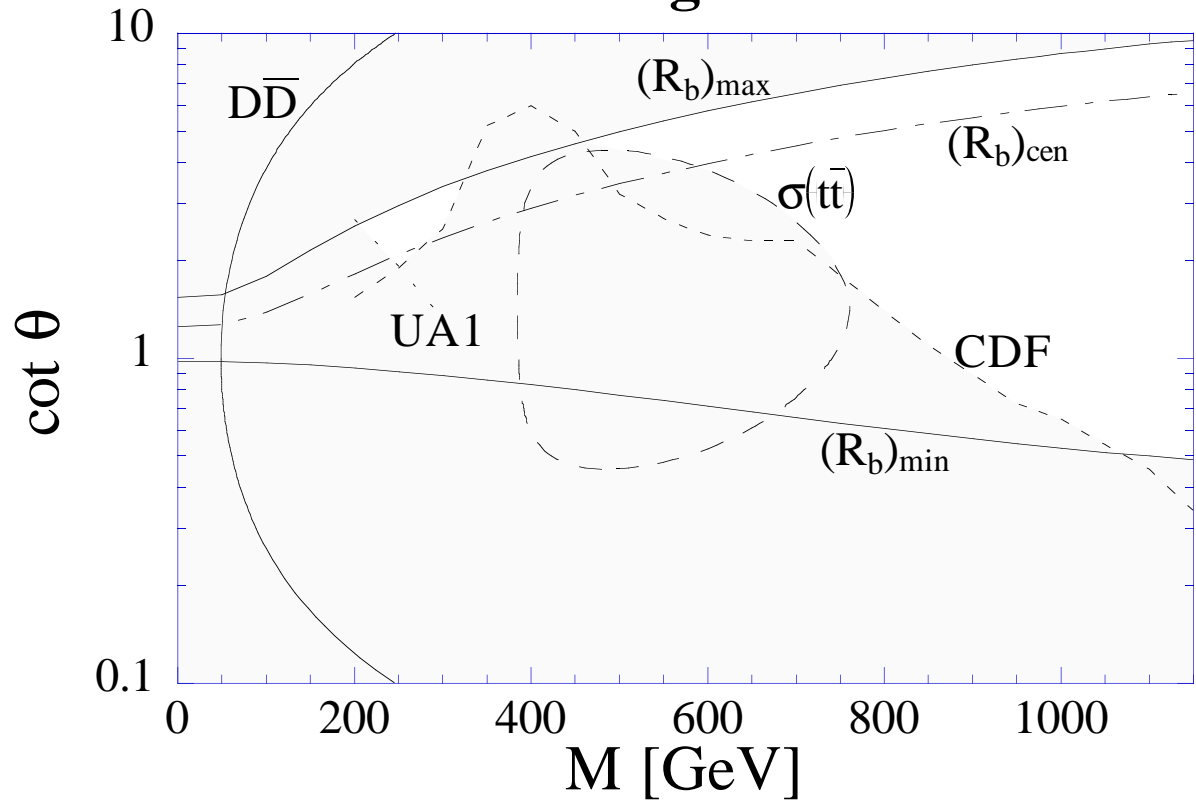
{\bf Fig. 1.}  Allowed regions of the $M$-$\cot\theta$ plane, for $\xi=0$. The
region above (below) the upper (lower) solid line is excluded by $R_b$, the
region below the dashed line is excluded by searches at UA1 and CDF for new
particles decaying to dijets, the region inside the dashed ``oval'' is excluded
by the $t\overline{t}$ production cross section, and the region to the left of
the solid parabola is excluded by $D$-$\overline{D}$ mixing.  The dotdashed
line shows the central value for $R_b$ measured at LEP.

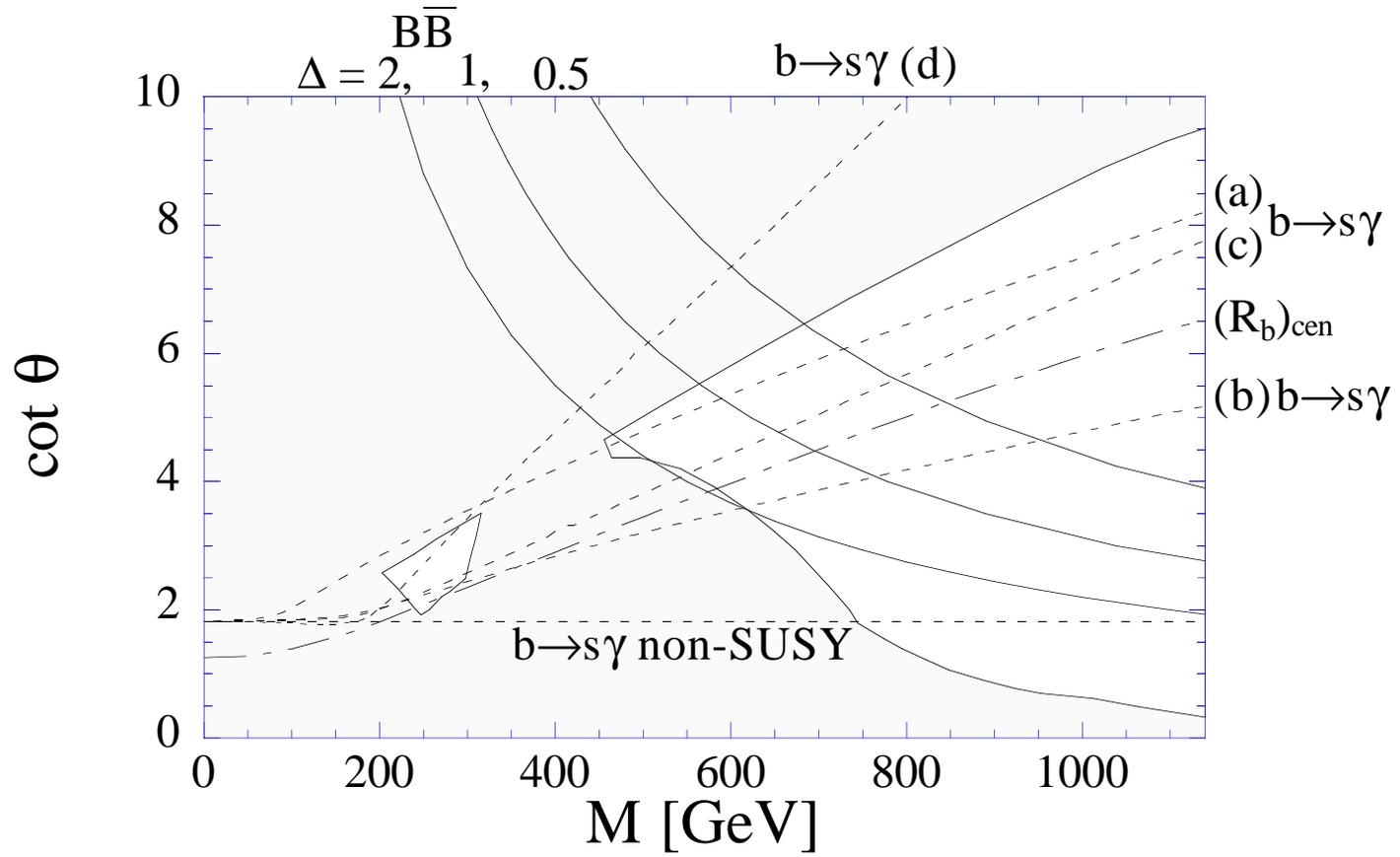
{\bf Fig. 2.} Allowed regions of the $M$-$\cot\theta$ plane, for
excluded by $B$-$\overline{B}$ mixing, for the choice of $\Delta$
shown.  The region above each dashed line is excluded by the
$b \rightarrow s \gamma$ constraint for $\Delta=1$, and
for various choices of the soft SUSY breaking masses ($m_{\tilde{b}_R}$,
$m_{\tilde{b}_L}$, $m_{\psi}$, $m_{\chi}$, $A$).  The dashed lines
correspond to the following mass sets (in GeV): (a) (100,100,100,100,0),
(b) (200,200,200,200,0), (c) (200,200,200,200,200),
(d) (300,100,200,200,200).  The non-supersymmetric result is
also shown.

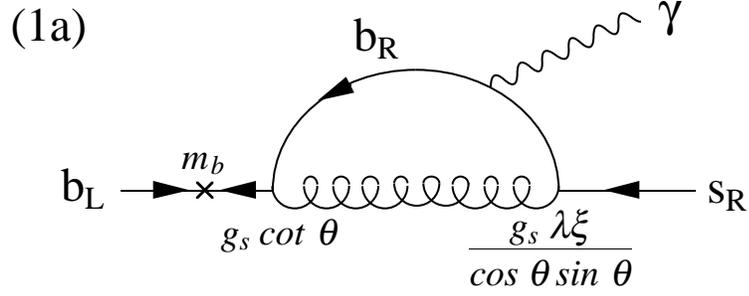
{\bf Fig. 3.} Feynman diagrams proportional to $1/\sin^2\theta$
that contribute to $b\rightarrow s \gamma$.  The numbering of the
diagrams corresponds to the results presented in Appendix~B.


\begin{thebibliography}{99}
\frenchspacing
\bibitem{pdg}  L.~Montanet {\it et al}\/., Particle Data Group, {\sl
Phys. Rev.}\/ {\bf D50}, 1173 (1994).
\bibitem{kag}
A.L.~Kagan, SLAC-PUB-6626, hep-ph/9409215.
\bibitem{shi}
M.~Shifman, TPI-MINN-94-42-T, hep-ph/9501222.
\bibitem{fn} C.D.~Froggatt and H.B.~Nielsen, {\sl Nucl. Phys.}\/ {\bf
B147}, 277 (1979).
\bibitem{early} A.~Salam, in {\sl Proceedings of the European Physical
Society International Conference on High Energy Physics},\/ Geneva,
1979, edited by A.~Zichichi (CERN, Geneva, 1980), footnote 41 therein;
S.~Rajpoot, {\sl Phys. Rev.}\/ {\bf D24}, 1890 (1981);
X.-Y.~Li and E.~Ma, {\sl Phys. Rev. Lett.}\/ {\bf 47}, 1788 (1981);
H.~Georgi, {\sl Nucl. Phys.}\/ {\bf B202}, 397 (1982);
\bibitem{ma}
M.~Fukugita, H.~Murayama, K.~Suehiro, and T.~Yanagida, {\sl Phys.
Lett.}\/ {\bf B283}, 142 (1992);
X.-Y.~Li and E.~Ma, {\sl Phys. Rev.}\/ {\bf D46}, 1905 (1992); {\sl J.
Phys.}\/ {\bf G19}, 1265 (1993).
\bibitem{hill} C.T.~Hill, {\sl Phys. Lett.}\/ {\bf B266}, 419 (1991);
C.T.~Hill and X.~Zhang, FERMILAB-PUB-94/231-T, hep-ph/9409315.
\bibitem{afar} L.F.~Abbott and E.~Farhi, {\sl Phys. Lett.}\/
{\bf B101}, 69 (1981); {\sl Nucl. Phys.}\/
{\bf B189}, 547 (1981);
\bibitem{laenen} E.~Laenen, J.~Smith and W.L.~van~Neerven,
{\sl Phys. Lett.}\/ {\bf B321}, 254 (1994).
\bibitem{cdf} CDF collabration, FERMILAB-PUB-95/022-E, hep-ex/9503002.
\bibitem{ua1} C.~Albajar {\it et al.}, UA1 collaboration, {\sl Phys.
Lett.}\/ {\bf B2-0}, 127 (1988).
\bibitem{cdf2} F.~Abe {\it et al.}, CDF collaboration, hep-ex/9501001.
Actually their earlier paper ({\it ibid.}\/, {\sl Phys. Rev. Lett.}\/
{\bf 71}, 2542 (1993)) quotes more stringent limits on the resonant
excess cross section, but they assume isotropic decays of the
resonance and hence an acceptance that is too optimistic.
\bibitem{amp} K.~Hagiwara and H.~Murayama, {\sl Phys. Lett.}\/ {\bf B246},
533 (1990).
\bibitem{soni} C.W.~Benard, J.N.~Labrenz, A.~Soni, {\sl Phys. Rev.}
\/{\bf D49}, 2536 (1994).
\bibitem{Buras} A.J. Buras, M. Jamin, and P.H. Weisz, {\sl Nucl. Phys.}
\/{\bf B347}, 491 (1990).
\bibitem{gsor} H.~Georgi, {\sl Phys. Lett.}\/ {\bf B297}, 353 (1992);
T.~Ohl, G.~Ricciardi, and E.H.~Simmons, {\sl Nucl. Phys.}\/ {\bf B403},
605 (1993);
\bibitem{glas} R.~Forty,CERN-PPE-94-154, Oct 1994, Review talk at Int.
Conf. on High Energy Physics, Glasgow, Scotland, Jul 20-27, 1994.
\bibitem{Altarelli} G.~Altarelli and S.~Petrarca, {\sl Phys. Lett.}\/
{\bf B261}, 303 (1991).
\bibitem{GSW} B.~Grinstein, R.~Springer and M.B.~Wise, {\sl Nucl.
Phys.}\/ {\bf B339}, 269 (1990).
\bibitem{muheim} F. Muheim, talk given at DPF '94,
Alberquerque, New Mexico, August, 1994.
\bibitem{susy}
S. Ferrara and E. Remiddi, {\sl Phys. Lett.}\/ {\bf B53}, 347 (1974);
R.~Barbieri and G.F.~Giudice, {\sl Phys. Lett.}\/ {\bf B309},
86 (1993)
\bibitem{CLEO} CLEO Collaboration (M.S. Alam, {\em et al.}) {\sl Phys. Rev.
Lett.}\/ {\bf 74}, 2885 (1995).
\bibitem{ciu} M. Ciuchini, {\em et al.},
{\sl Phys. Lett.}\/ {\bf B334}, 137 (1994)
\bibitem{compl}
G. 't Hooft, Carg$\grave{{\rm e}}$se Summer Institute Lectures (1979).
\end{thebibliography}
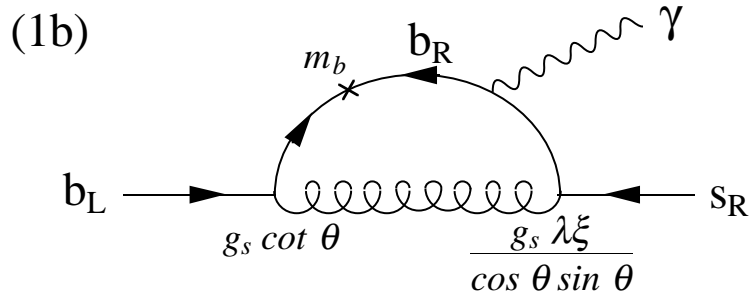
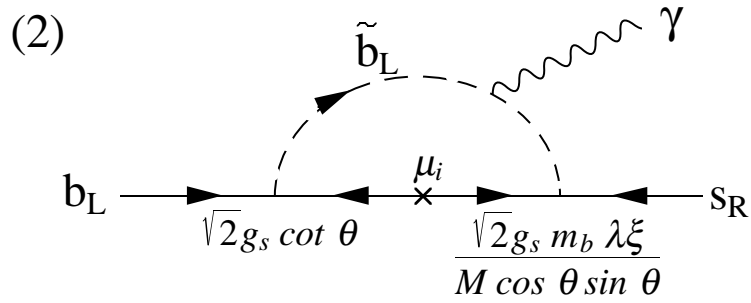
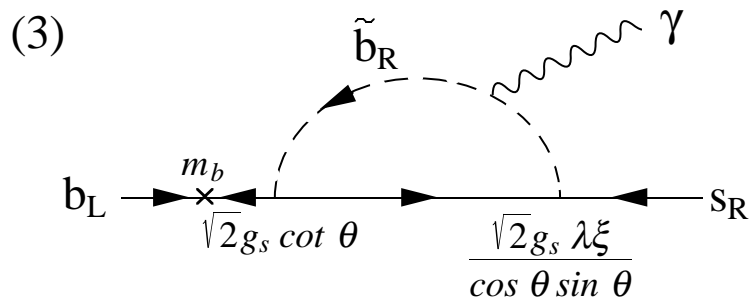
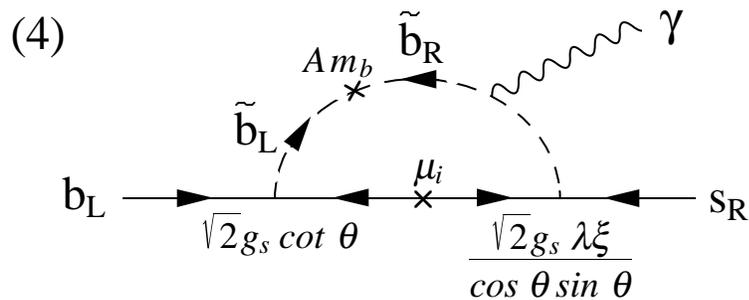
\end{document}